\documentclass[final,3p,times,twocolumn]{elsarticle}
\usepackage{amsmath}
\usepackage{amssymb}
\usepackage{epstopdf}
\usepackage[T1]{fontenc}
\usepackage{lineno,hyperref}
\modulolinenumbers[5]
\usepackage{subcaption}
\usepackage{xcolor}
\usepackage{arydshln}
\usepackage{soul}
\usepackage{comment}
\setstcolor{red}

\usepackage{hyperref}

\journal{Journal of Quantitative Spectroscopy and Radiative Transfer}

\begin{document}

\begin{frontmatter}

\title{\textit{FitAik}: a package to calculate least-squares fitted atomic transitions probabilities. Application to the Er$^+$ lanthanide ion}

\author[1]{Maxence Lepers}

\affiliation[1]{organization={Laboratoire Interdisciplinaire Carnot de Bourgogne, UMR 6303, CNRS, Univ.~Bourgogne Franche-Comte},
                addressline={9 Avenue Alain Savary, BP 47870}, 
                postcode={F-21078},
                city={Dijon cedex},
                country={France}}

\author[2]{Olivier Dulieu}
\author[2]{Jean-Fran{\c c}ois Wyart}
\affiliation[2]{organization={Laboratoire Aime Cotton, CNRS, Universite Paris-Saclay},
                addressline={B{\^ a}timent 505},
                city={Orsay},
                postcode={F-91405}, 
                country={France}}

\begin{abstract}
We present a new method implemented in our new package \textit{FitAik}, to perform least-squares fitting of calculated and experimental atomic transition probabilities, by using the mono-electronic transition integrals $\langle n\ell |r| n'\ell' \rangle$ (with $r$ the electronic radial coordinate) as adjustable quantities. \textit{FitAik} is interfaced to the Cowan suite of codes, for which it automatically writes input files and reads output files. We illustrate our procedure with the example of Er$^{+}$ ion, for which the agreement between calculated and experimental Einstein coefficients is found to be very good. The source code of \emph{FitAik} can be found on GitLab, and the calculated Einstein coefficients are stored in our new database CaDDiACs. They are also used to calculate the dynamic dipole polarizability of Er$^+$.
\end{abstract}


\end{frontmatter}

\section{Introduction}

The spectroscopy of lanthanide ions has long been studied in the context of astrophysics, as shown by the number of articles published on that topic in astrophysical journals, see \textit{i.e.} \cite{biemont2000, biemont2002, zhang2002, lawler2008, wyart2008, tian2019, radvziute2020, radvziute2021}. As examples of interest, one can cite the study of chemically-peculiar stars \cite{cowley1978, cowley1984, ryabchikova2006}, or the so-called r-process in neutron star mergers \cite{kasen2013, fontes2015, watson2019}.

\begin{figure}[h]
  \includegraphics[width=0.5\textwidth]{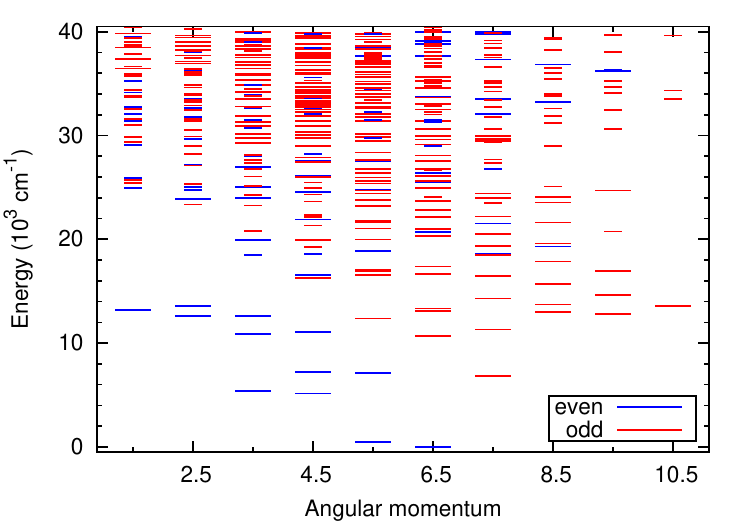}
  \caption{Energy levels of Er$^+$ sorted according to the total electronic angular momentum $J$ and the parity (even parity in blue, odd parity in red). The long lines correspond to experimental energies, while the short ones to calculated energies of experimentally unknown levels.
  \label{fig:lev}}
\end{figure}

In a different context, the spectroscopy of Rydberg states of erbium has recently been investigated experimentally \cite{trautmann2021}, following an earlier study on holmium \cite{hostetter2015}. Both groups are involved in the development of experiments with ultracold gases of lanthanide atoms \cite{aikawa2012, miao2014, frisch2015, trautmann2018} that has taken place for 15 years \cite{norcia2021, chomaz2022}. Rydberg atoms with several valence electrons offer the possibility to use their open-shell ionic core for \textit{e.g.}~laser cooling or trapping, based on isolated-core excitation, see \textit{i.e.} \cite{lehec2021, pham2022, burgers2022}. Yet those purposes require a precise knowledge of the core energies, transition intensities and dynamic polarizabilities. In this respect, we investigated in 2016 candidates for laser-cooling transitions in Er$^{+}$ \cite{lepers2016}, relying on an accurate modeling of the Er$^{+}$ spectrum (see Fig.~\ref{fig:lev}), whose description motivates the present article.

To perform such atomic-structure calculations, Robert D.~Cowan's suite of codes is a widely used tool for more than forty years \cite{cowan1981, kramida2019}. It consists of four Fortran programs called \textit{RCN}, \textit{RCN2}, \textit{RCG} and \textit{RCE}, which can be downloaded on the website of the University of Dublin \cite{cowan-dublin}. Based on the same architecture, A.~Kramida wrote his own version of the codes, improving the performance of the original ones and correcting some major bugs \cite{kramida2019}. Those two versions contain the so-called \textit{RCE} program to perform a least-squares fitting of experimental energies and Hamiltonian eigenvalues calculated \textit{ab initio}. However, the mono-electronic transition integrals that are the building blocks of the transition intensities have their \textit{ab initio} values. In parallel, P.~Quinet and coworkers have modified the \textit{ab initio} part of the Cowan codes, in order to account for core-polarization effects in the calculation of single electron wave functions \cite{quinet1999, quinet2002}. This has significantly increased the accuracy of the predicted intensities (Einstein coefficients or oscillator strengths), in comparison with the \textit{ab initio} ones of the original codes.

Following the idea of least-squares fitting of energies, we have developed a suite of codes called \textit{FitAik}, that we interfaced to Cowan codes, to perform a least-squares fitting of transition probabilities, \textit{i.e.}~Einstein coefficients of spontaneous emission, by considering mono-electronic transition integrals $\langle n\ell |r| n'\ell' \rangle$ (where $r$ is the electronic radial coordinate) as variable quantities. Our method allows for accurately reproducing many measured Einstein coefficients, and for predicting yet unknown ones. In this article, we describe in detail our fitting procedure, and we illustrate it with Er$^{+}$ \cite{lepers2016}, for which J.~E.~Lawler's group has provided a set of approximately 400 experimental Einstein coefficients \cite{lawler2008}. The agreement between calculated and experimental coefficients is found to be very satisfactory. 
Following a similar semi-empirical methodology, J.~Ruchkowski and coworkers developed their own numerical code to fit energies, hyperfine constants, and oscillator strengths, which they applied \textit{e.g.}~to scandium ion Sc$^{+}$ \cite{ruczkowski2014} and strontium atom Sr \cite{ruczkowski2016}. To the best of our knowledge, this code is not open-source.

The article is organized as follows: Section \ref{sec:th} presents the theoretical background of our method, while Section \ref{sec:res} presents our results in the case of Er$^{+}$, and Section \ref{sec:ccl} contains concluding remarks.

\section{Theoretical background}
\label{sec:th}

In this section, we present the theoretical basis of our calculations of Einstein coefficients for atomic transitions, illustrated, for the sake of clarity, with the example of the Er$^+$ ion, for which the results will be given in Section \ref{sec:res}.

\subsection{The expression of Einstein coefficients}

We consider a spontaneous emission (SE) transition from an upper level $|i\rangle$ of energy $E_i$ and total electronic angular momentum $J$, to a lower level $|k\rangle$ of energy $E_k$ and total electronic angular momentum $J'$. The corresponding Einstein coefficient for SE is given by
\begin{equation}
  A_{ik} = \frac{\omega_{ik}^{3}}{3\pi\varepsilon_0 \hbar c^3 (2J+1)}
           \left| \left\langle i \left\| \mathbf{d}
           \right\| k \right\rangle \right|^2
  \label{eq:aik}
\end{equation}
where $\varepsilon_0$ is the vacuum permittivity, $\hbar$ the reduced Planck constant, $c$ the speed of light and $\omega_{ik} = 2\pi \nu_{ik} = (E_i-E_k)/\hbar$ the transition frequency. The quantity $\langle i\| \mathbf{d} \| k \rangle$ is the reduced matrix element of the electric dipole moment (EDM), equal to $\mathbf{d} = -e \sum_\alpha \mathbf{r}_{\alpha}$. where $e$ is the elementary charge and $\mathbf{r}_{\alpha}$ the instantaneous position of the $\alpha$-th electron.

In practice, the eigenvectors describing each atomic level are expanded on a basis set written in the framework of the Russel-Saunders ($LS$) coupling
\begin{equation}
  \left|i \right\rangle = \sum_{b}  c_{b}  \left|b , J \right\rangle
  \,\, \mathrm{and} \,\,
  \left|k \right\rangle = \sum_{b'} c_{b'} \left|b', J'\right\rangle
  \label{eq:bas}
\end{equation}
A given basis state $|b, J\rangle$ consists of an electronic configuration and of intermediate orbital and spin angular momenta, and it has a well-defined electronic parity, odd or even.

Table \ref{tab:config} presents the electronic configurations included in our Er$^+$ calculations: there are four configurations of even parity and five configurations of odd parity. For each configuration, the table also presents the number of basis states for $J=13/2$.
For instance, in the case of $4f^{12}6s$, the two possible $LS$ states are $4f^{12}({}^3H)6s(^2S) \,{}^4H$ and $4f^{12}({}^1I)6s(^2S) \,{}^2I$, where the spectral terms in parentheses refer to individual subshells $4f^{12}$ and $6s$, and the one without parentheses gives the total orbital and spin angular momenta of the states. As another example, the five possible ones for $4f^{12}6p$ are: $4f^{12}({}^3H)6p({}^2P^{\text{o}}) \,{}^4I^{\text{o}}$, ${}^4H^{\text{o}}$, or ${}^2I^{\text{o}}$, and $4f^{12}({}^1I)6p({}^2P^{\text{o}}) \, {}^2K^{\text{o}}$ or ${}^2I^{\text{o}}$. Note that for $4f^{13}$, the only possible $LS$ term is ${}^2F^{\text{o}}$, possessing two states with $J=5/2$ and $7/2$, but not with $13/2$.

\begin{table}[h]
  \caption{Electronic configurations sorted by even/odd parity included in our Er$^+$ (Er II) calculations, as well as the number of basis states for $J=13/2$. 
  }
  \label{tab:config}
  \begin{center}
  \begin{tabular}{cc|cc}
  \hline
    \multicolumn{2}{c|}{Even parity} & \multicolumn{2}{c}
    {$\phantom{I^{I^I}}$Odd parity$\phantom{I^{I^I}}$} \\
    Config. & Nb.~states & Config. & Nb.~states \\
  \hline
    $4f^{12}6s$   & $\phantom{I^{I^I}}$2$\phantom{I^{I^I}}$ &
    $4f^{13}$     & 0 \\
    $4f^{12}5d$   & 9 &
    $4f^{11}6s^2$ & 3 \\
    $4f^{11}6s6p$ & 44 &
    $4f^{11}5d6s$ & 75 \\
    $4f^{11}5d6p$ & 223 &
    $4f^{11}5d^2$ & 166 \\
     & &
    $4f^{12}6p$   & 5 \\
  \hline
  \end{tabular}
  \end{center}
\end{table}

By introducing Eq.~\eqref{eq:bas} in the reduced EDM of Eq.~\eqref{eq:aik}, we get
\begin{equation}
  \left\langle i \left\| \mathbf{d} \right\| k \right\rangle
    = \sum_{bb'} c_{b} c_{b'} \left\langle b,J \left\| 
      \mathbf{d} \right\| b',J' \right\rangle .
  \label{eq:tdm}
\end{equation}
To yield a non-zero contribution, a $(b,b')$ pair must involve configurations that differ by only one electron, \textit{e.g.}~$4f^{12}6s$-$4f^{12}6p$ or $4f^{12}6s$-$4f^{11}5d6s$. The subshells of the {}``hopping'' electron, labeled $(n\ell, n'\ell')$, must also satisfy $\ell'-\ell = \pm 1$. Among the $4\times 5=20$ pairs of configurations with opposite parities, 10 obey those selection rules (see Table \ref{tab:rj}), and the corresponding EDM matrix element $\langle b,J \| \mathbf{d} \| b',J' \rangle$ is proportional to the one-electron position operator $r_{n\ell,n'\ell'} = \langle n\ell |r| n'\ell' \rangle$. As a consequence, all the EDM matrix elements depend only on ten $r_{n\ell,n'\ell'}$ quantities, as shown in Table \ref{tab:rj}. We can thus rewrite Eq.~\eqref{eq:aik} in the general form
\begin{equation}
  A_{t} = \left( \sum_{j=1}^{N_\mathrm{par}} \overline{a}_{tj} \, r_j \right)^2
  \label{eq:aik-2}
\end{equation}
where $t \equiv (ik)$ is an index characterizing the transition between $|i\rangle$ and $|k\rangle$, and $j$ the $N_\mathrm{par}$ possible pairs of subshells $(n\ell, n'\ell')$. The quantities $\overline{a}_{tj}$ depend on the transition frequency $\nu_{ik}$, the coefficients ($c_{b}, c_{b'}$), and on the angular momenta of the states in a complex way (see Ref.~\cite{cowan1981}, Chap.~14). In what follows, the quantities $r_j$ will be treated as adjustable parameters.

\begin{table}
  \caption{Pairs of opposite parity configurations and subshells $(n\ell, n'\ell')$ obeying the electric-dipole selection rule, as well as the corresponding integral $r_j \equiv r_{n\ell,n'\ell'}$ calculated by the Hartree-Fock + relativistic (HFR) method, and the group of free parameters to which they belong (see Subsection \ref{sub:aik}).
    \label{tab:rj}}
  \begin{center}
  \begin{tabular}{cccrc}
  \hline
    Even conf. & Odd conf. & $(n\ell, n'\ell')$ & $r_{n\ell,n'\ell'}$ 
    & Group \\
  \hline
     $\phantom{I^{I^I}}4f^{12}6s\phantom{I^{I^I}}$  
                   & $4f^{11}5d6s$ & $4f$-$5d$ &  0.5171 & 5 \\
     $4f^{12}6s$   & $4f^{12}6p$   & $6s$-$6p$ & -3.4883 & 1 \\
     $4f^{12}5d$   & $4f^{13}$     & $5d$-$4f$ &  0.6322 & 5 \\
     $4f^{12}5d$   & $4f^{11}5d^2$ & $4f$-$5d$ &  0.5166 & 5 \\
     $4f^{12}5d$   & $4f^{12}6p$   & $5d$-$6p$ &  2.4909 & 4 \\
     $4f^{11}6s6p$ & $4f^{11}6s^2$ & $6p$-$6s$ & -3.3745 & 2 \\
     $4f^{11}6s6p$ & $4f^{11}5d6s$ & $6p$-$5d$ &  1.8709 & 4 \\
     $4f^{11}5d6p$ & $4f^{11}5d6s$ & $6p$-$6s$ & -3.0151 & 3 \\
     $4f^{11}5d6p$ & $4f^{11}5d^2$ & $6p$-$5d$ &  2.0789 & 4 \\
     $4f^{11}5d6p$ & $4f^{12}6p$   & $5d$-$4f$ &  0.5188 & 5 \\
  \hline
  \end{tabular}
  \end{center}
\end{table}

\subsection{The least-squares fitting procedure}

We use a set of $N_\mathrm{tr}$ experimental Einstein coefficients $A_{t,\mathrm{exp}}$, $t \in [1; N_\mathrm{tr}]$, published by the Wisconsin group \cite{lawler2008} in the case of Er$^+$. In our least-squares fitting procedure, we seek to minimize the standard deviation $\sigma_{A}$,
\begin{equation}
  \sigma_{A} = \left[ \frac{ \sum_{i=1}^{N_\mathrm{tr}} 
    \left( A_{t,\mathrm{cal}} - A_{t,\mathrm{exp}} \right)^2}
    {N_\mathrm{tr}-N_\mathrm{par}} \right]^{1/2},
  \label{eq:stdev-a}
\end{equation}
where $A_{t,\mathrm{cal}}$ is given by Eq.~\eqref{eq:aik-2}. Because the Einstein coefficients can be spread over several orders of magnitude, minimizing Eq.~\eqref{eq:stdev-a} may tend to minimize in priority the error on the strongest transitions. To avoid this, we also define the logarithmic standard deviation  $\sigma_{\mathrm{log} A}$
\begin{equation}
  \sigma_{\mathrm{log}A} = \left[ \dfrac{ \sum_{i=1}^{N_\mathrm{tr}} 
    \log^2\left( \dfrac{A_{t,\mathrm{cal}}} {A_{t,\mathrm{exp}}} \right)}
    {N_\mathrm{tr}-N_\mathrm{par}} \right]^{1/2} .
  \label{eq:stdev-lga}
\end{equation}
For both quantities, a first calculation is performed on a grid of discrete $r_j$ parameters (see below). Defining
\begin{equation}
  r_j = f_j \, r_{j,\mathrm{init}},
\end{equation}
with $f_j$ the ratio, or scaling factor (SF), between the $r_j$ variables and their initial values $r_{j,\mathrm{init}}$, namely the Hartree-Fock + relativistic (HFR) ones calculated by the Cowan code RCN2. Equation~\eqref{eq:aik-2} becomes
\begin{equation}
  A_{t} = \left( \sum_{j=1}^{N_\mathrm{par}} a_{tj} \, f_j \right)^2,
  \label{eq:aik-3}
\end{equation}
where $a_{tj} = \overline{a}_{tj} \, r_{j,\mathrm{init}}$. Similarly to the energy least-squares fitting in the \textit{RCE} program of the Cowan suite, it is possible to force certain parameters to have the same SF during the calculation. This is, for instance, the case for all $r_{4f,5d}$ parameters of Table \ref{tab:rj}. The grid of the first least-squares fit is defined on SFs: their minimum $f_{j,\mathrm{min}}$, maximum $f_{j,\mathrm{max}}$ and step $\delta f_{j}$.

In a second step, the optimal grid for SFs serves as the set of initial values for a more precise fit based on the Gauss-Newton method. As the Einstein coefficients are non-linear functions of the fitting parameters $r_j$ (or $f_j$), the success of this second fit requires those initial values to be reasonably close to the final solution. Namely, we search for the set of parameters $f_j$ gathered in the vector $\mathbf{F}$ for which the gradient of the standard deviation $\mathbf{\nabla}_\mathbf{F} \sigma_A = \mathbf{0}$. The components of the gradient vector are given by
\begin{equation}
  \frac{\partial \sigma_A}{\partial f_j} = \frac{2}
    {(N_\mathrm{tr}-N_\mathrm{par}) \, \sigma_A}
    \sum_{t=1}^{N_\mathrm{tr}} a_{tj} 
    (A_{t,\mathrm{cal}} - A_{t,\mathrm{exp}})
    \sqrt{A_{t,\mathrm{cal}}}.
  \label{eq:grad}
\end{equation}
Given the initial set of parameters $\mathbf{F}_0$ (resulting from the grid calculation), the set obtained after the first iteration $\mathbf{F}_1$ is then equal to
\begin{equation}
  \mathbf{F}_1 = \mathbf{F}_0 - \left(\mathbf{J}_{\mathbf{F}_0}\right)^{-1} 
    \, \mathbf{\nabla}_{\mathbf{F}_0},
  \label{eq:F1}
\end{equation}
where $\mathbf{J}_\mathbf{F}$ is the $N_\mathrm{par} \times N_\mathrm{par}$ Jacobian matrix, whose elements are equal to
\begin{align}
  J_{ij} & = \frac{\partial^2 \sigma_A}{\partial f_i \, \partial f_j}
    \nonumber \\
   & = \frac{2}{(N_\mathrm{tr}-N_\mathrm{par}) \, \sigma_A}
    \sum_{t=1}^{N_\mathrm{tr}} a_{ti} a_{tj}
    (3A_{t,\mathrm{cal}} - A_{t,\mathrm{exp}})
  \nonumber \\
   & - \frac{4}{(N_\mathrm{tr}-N_\mathrm{par})^2 \, \sigma_A^3}
    \sum_{t=1}^{N_\mathrm{tr}} a_{ti}
    (A_{t,\mathrm{cal}} - A_{t,\mathrm{exp}}) \sqrt{A_{t,\mathrm{cal}}}
  \nonumber \\
   & \quad \times \sum_{u=1}^{N_\mathrm{tr}} a_{uj}
    (A_{u,\mathrm{cal}} - A_{u,\mathrm{exp}})
    \sqrt{A_{u,\mathrm{cal}}}.
  \label{eq:jacob}
\end{align}
Equation \eqref{eq:F1} is repeated until convergence is reached, namely $|\partial \sigma_A / \partial f_j| \le \epsilon$, $\forall j$, where $\epsilon$ is arbitrarily small.

Once the convergence is reached, we estimate the root-mean-square deviation $\Delta f_j$ of a given SF, by assuming that, when the SF varies by $\pm \Delta f_j$ around its optimal value $f_{j,\mathrm{opt}}$, the standard deviation $\sigma_A$ increases by the quantity $\delta$ (say 5~\% which is a typical uncertainty of the Wisconsin group's measurements). Using a second-order Taylor expansion around the optimal SFs, one obtains 
\begin{equation}
  \Delta f_j = \sqrt{\frac{2\delta\, \sigma_{A,\mathrm{opt}}}
               {J_{jj,\mathrm{opt}}}},
  \label{eq:sensit}
\end{equation}
where $\sigma_{A,\mathrm{opt}}$ and $J_{jj,\mathrm{opt}}$ are respectively the standard deviation and diagonal Jacobian matrix elements obtained with the optimal set of SFs.

The same iterative approach as in Eq.~\eqref{eq:F1} can be followed with the logarithmic standard deviation \eqref{eq:stdev-lga}, except that the gradient vector and Jacobian matrix will have slightly different expressions. In order to avoid the repetition of long equations, the latter are given in \ref{sec:stdev-log}.

\section{Results for Er$^+$}
\label{sec:res}

This section is dedicated to the calculations of Einstein coefficients. For the sake of completeness we first briefly discuss the calculated level energies.

\subsection{Energy levels}

The modeling of level energies of Er$^+$ was the purpose of Ref.~\cite{wyart2009}, in which the authors gave optimal sets of energy parameters for both parities. Respectively, four and five electronic configurations were considered in the even and odd parities (Table \ref{tab:config}). The odd configuration $4f^{13}$ is included for a technical purpose regarding the Cowan codes, but no experimental level belonging to it was observed. In the even parity, 130 levels were fitted with 25 free parameters, giving a standard deviation of 55~cm$^{-1}$; in the odd parity, 233 levels were fitted with 21 free parameters, giving a standard deviation of 63~cm$^{-1}$. Figure \ref{fig:lev} shows experimental energies when they have been detected \cite{NIST_ASD}, and calculated ones otherwise, as functions of the electronic angular momentum $J$, sorted by parity.

\begin{table*}
  \caption{Odd-parity levels added in the present version of the fit: $E_\mathrm{exp}$, $E_\mathrm{cal}$ stand for experimental and calculated energies in cm$^{-1}$, $g_\mathrm{exp}$ and $g_\mathrm{cal}$ for Land\'e $g$-factors. The last level was included in Ref.~ \cite{wyart2009}, but no eigenvector was specified.
  \label{tab:adlev}}
  \begin{center}
  \begin{tabular}{rrcrrrr}
  \hline
   $E_\mathrm{exp}$ & $E_\mathrm{cal}$ & $J$ & $g_\mathrm{exp}$ &
    $g_\mathrm{cal}$ & \multicolumn{2}{c}{Leading term \& percentage} \\
  \hline
   39053.059 & 39002 & $7/2$ & 0.980 & 1.031 &
     $4f^{11}(^4F_{5/2}^o)5d6s(^1D_2) \,\, (5/2,2)^{\text{o}}$ & 17.0 \\
   42527.301 & 42488 & $7/2$ & 1.020 & 1.190 & 
     $4f^{11}(^4S_{3/2}^o)5d^2(^3F_2) \,\, (3/2,2)^{\text{o}}$ & 11.3 \\
   43221.645 & 43305 & $9/2$ & 1.060 & 1.075 &
     $4f^{11}(^2G_{7/2}^o)5d6s(^3D_3) \,\, (3/2,3)^{\text{o}}$ & 8.5 \\  
   44148.047 & 44219 & $7/2$ & 1.065 & 1.041 & 
     $4f^{11}(^4I_{11/2}^o)5d^2(^1G_4) \,\, (11/2,4)^{\text{o}}$ & 7.8 \\
   44162.145 & 44199 & $3/2$ & 0.770 & 0.893 & 
     $4f^{11}(^4I_{11/2}^o)5d^2(^1G_4) \,\, (11/2,4)^{\text{o}}$ & 17.9 \\
  \hline
  \end{tabular}
  \end{center}
\end{table*}

The agreement between the calculated and experimental energies is very satisfactory, since the standard deviations in the two parities are similar to those obtained in other lanthanide ions with our semi-empirical method \cite{wyart2011}. Moreover, a recent purely \textit{ab initio} calculation reports on a relative average deviation of 4~\%, which in regard to the energy range of about 40000~cm$^{-1}$ covered by the calculation, corresponds to an absolute average deviation of about 1600~cm$^{-1}$ (see Ref.~\cite{radvziute2021}, Table 11). Compared to Ref.~\cite{wyart2009}, several odd-parity levels listed in Table \ref{tab:adlev} previously excluded are introduced in the present fit. This does not significantly change the optimal energy parameters, that are given in the RCG input files {}``Er+\_opt.ing11'' and {}``Er+\_opt.ING11'' in the Supplementary Material.

\subsection{Einstein coefficients}
\label{sub:aik}

Once the fitting of level energies is done, we tackle the fitting of Einstein coefficients, for which we use the experimental set of data given in Ref.~\cite{lawler2008}, containing 418 transitions. We aim at adjusting the SFs $f_j$ given in Table \ref{tab:rj}. Firstly, we determine which groups of $f_j$ are forced to remain equal during the fitting process. If we let all 10 parameters to vary freely, we sometimes obtain non-physical optimal values (\textit{i.e.} much larger than one), especially $f_{5d,4f}$ for the $4f^{12}5d$-$4f^{13}$ transitions, since $4f^{13}$ possess no experimental levels. After trying different types of constraints, we obtain the most satisfactory results by forcing all the $f_{6p,5d}$ and all the $f_{5d,4f}$ parameters to be equal to each other, while the three $f_{6s,6p}$ evolve freely. This yields the five groups of free parameters given in Table \ref{tab:rj}.

\begin{table}
  \caption{Optimal scaling factors $f_j$ with their uncertainties $\Delta f_j$ given by Eq.~\eqref{eq:sensit}, as well as the linear $\sigma_A$ (in s$^{-1}$),  and logarithmic  $\sigma_{\mathrm{log}A}$ standard deviations (Eqs.~\eqref{eq:stdev-a} and \eqref{eq:stdev-lga}), obtained in different cases: (1) With all the experimental lines of Ref.~\cite{lawler2008}; (2) Excluding the lines given in Table \ref{tab:excl-trs}, which show a large discrepancy between the calculated and experimental Einstein coefficients; (3) Same data as (2) but minimizing the logarithmic \eqref{eq:grad-lg} rather than the linear standard deviation \eqref{eq:stdev-a}.
  \label{tab:opt-sf}}
  \begin{center}
  \begin{tabular}{cccc}
  \hline
   $\phantom{{}^{|^|}}$SF$\phantom{{}^{|^|}}$ & (1) & (2) & (3) \\
  \hline
   $f_1$ & $0.884 \pm 0.056$ & $0.886 \pm 0.046$ & $0.987 \pm 0.081$ \\
   $f_2$ & $0.877 \pm 0.055$ & $0.876 \pm 0.044$ & $0.892 \pm 0.607$ \\
   $f_3$ & $0.797 \pm 0.088$ & $0.797 \pm 0.071$ & $0.870 \pm 0.187$ \\
   $f_4$ & $0.799 \pm 0.493$ & $0.808 \pm 0.394$ & $0.857 \pm 0.099$ \\
   $f_5$ & $0.822 \pm 0.701$ & $0.817 \pm 0.569$ & $0.859 \pm 0.179$ \\
  \hline
   $\sigma_A$ & ${}^{\phantom{I^I}} 5.5 \times 10^6 {}^{\phantom{I^I}}$ 
   & $4.6 \times 10^6$ & $5.9 \times 10^6$ \\
   $\sigma_{\mathrm{log}A}$ & 0.52 & 0.22 & 0.20 \\
  \hline
  \end{tabular}
  \end{center}
\end{table}

\begin{figure}
  \includegraphics[width=0.5\textwidth]{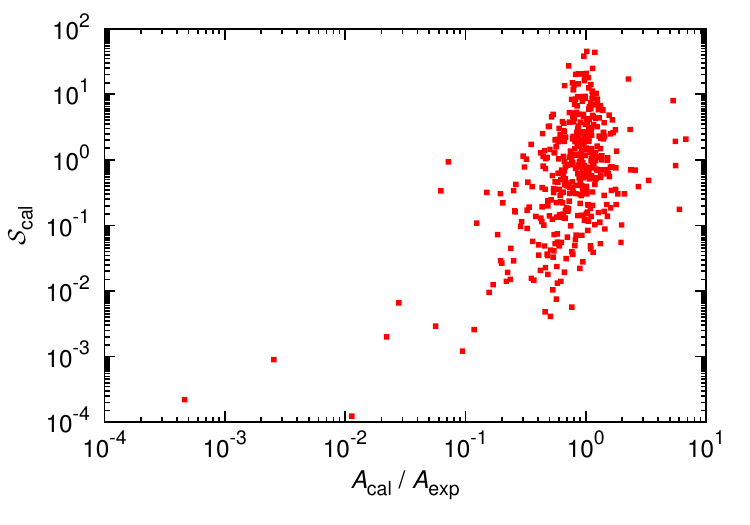}
  \caption{Calculated line strength $\mathcal{S}_\mathrm{cal}$ (in atomic units) as function of the ratio between calculated and experimental Einstein coefficients in log scale, obtained with the optimal scaling factors of column ``(1)'' in Table \ref{tab:opt-sf}, \textit{i.e.}~including all the experimental transitions of Ref.~\cite{lawler2008}.
  \label{fig:ratio-scal}}
\end{figure}

In that case, we make a fit with all the experimental lines \cite{lawler2008}, which converges to the optimal SFs and uncertainties given in column {}``(1)'' of Table \ref{tab:opt-sf}. The corresponding linear standard deviation is $\sigma_A = 5.49 \times 10^6$~s$^{-1}$, which is 2.7~\% of the largest experimental Einstein coefficient. The logarithmic one is $\sigma_{\mathrm{log}A} = 0.52$, meaning that a majority of the ratios $A_{t,\mathrm{cal}} / A_{t,\mathrm{exp}}$ are larger than $10^{-0.52} \approx 0.30$ and smaller than $10^{0.52} \approx 3.3$. To visualize how accurately each experimental transition is reproduced, 
we plot on Figure \ref{fig:ratio-scal} the calculated line strength $\mathcal{S}_\mathrm{cal} = |\langle i\|\mathbf{d}\| k\rangle|^2$ as a function of the ratio $A_{\mathrm{cal}} / A_{\mathrm{exp}}$, both in logarithmic scale.

One can see that most ratios have values around one, even though a few ones are very small, down to $4.6 \times 10^{-4}$.
The transitions characterized by very small ratios are associated with very small calculated line strengths. Namely, the six transitions with a ratio smaller than 0.06 have line strengths smaller than 0.007 atomic units. This relationship has been pointed out in Ref.~\cite{kramida2013}. Therefore, $\mathcal{S}_\mathrm{cal}$ can be a suitable criterion to evaluate the reliability of calculated Einstein coefficients, in particular for those which have no experimental counterpart.

On the other hand, Figure \ref{fig:ratio-scal} also displays some transitions with a large $A_{\mathrm{cal}} / A_{\mathrm{exp}}$ ratio, the largest one being 6.81. To build the data set of calculation ``(2)'' from ``(1)'' of Table \ref{tab:opt-sf}, we exclude the transitions for which $A_{\mathrm{cal}} / A_{\mathrm{exp}} < 0.2$ or $A_{\mathrm{cal}} / A_{\mathrm{exp}} > 5$; those transitions are reported in Table \ref{tab:excl-trs}. Most of them involve upper levels of odd parity from 30000 to 40000~cm$^{-1}$.
In that range, the large density of levels can result in pairs of very close levels (less than 100~cm$^{-1}$ apart) with the same angular momentum. A list of such pairs is given in Table \ref{tab:lev-cls}.
Among them, transitions implying levels of the first two pairs show satisfactory agreement between calculated end experimental Einstein coefficients. The accuracy in the third pair is less satisfactory, \textit{e.g.}~the 7th line of Table \ref{tab:excl-trs} has $A_\mathrm{cal} / A_\mathrm{exp} = 5.5$; but inverting the two levels does not significantly improve it. For the pair at 37098 and 37147~cm$^{-1}$, inverting the two levels improves the fit with all lines, whose standard deviation drops from 5.5 to $4.6 \times 10^6 \,s^{-1}$. Still, it is worthwhile noting that the optimal scaling factors almost do not change. For the other pairs, there are no experimental Einstein coefficient to compare with. However, for the last level pair, the experimental and theoretical values of the Landé $g$-factors indicate a probable level inversion.

Excluding the lines of Table \ref{tab:excl-trs} defines calculation {}``(2)'' in Table \ref{tab:opt-sf}, for which we obtain a minimal standard deviation of $\sigma_A = 4.565 \times 10^6$\,s$^{-1}$. Those SFs yield a logarithmic standard deviation $\sigma_{\mathrm{log}A} = 0.217$, which represents a significant improvement compared to the fit with the full experimental spectrum for which $\sigma_A = 5.49 \times 10^6$~s$^{-1}$ and $\sigma_{\mathrm{log}A} = 0.524$. With the same set of experimental data, we also seek the SFs minimizing the logarithmic standard deviation, see calculation {}``(3)''. We obtain $\sigma_A = 5.891 \times 10^6$\,s$^{-1}$ and $\sigma_{\mathrm{log}A} = 0.199$, which means that for the majority of the transitions the ratio $ A_{\mathrm{cal}} / A_{\mathrm{exp}}$ is between $10^{-0.199} = 0.631$ and $10^{0.199} = 1.58$.

Even if some optimal SFs, like $f_1$ and $f_3$ differ notably in calculations {}``(2)'' and {}``(3)'', their ranges of uncertainty always overlap, namely, $f_1 = 0.886 \pm 0.046$ for set (2) and $0.987 \pm 0.081$ for set (3). In order to determine which set of SFs is the most suitable, we notice the following. In calculation {}``(2)'', the SFs minimizing $\sigma_A$ yield a $\sigma_{\mathrm{log}A}$ that is 9~\% larger than the lowest one given by calculation {}``(3)''. On the contrary, the $\sigma_A$ obtained in {}``(3)'' is 29~\% larger than the minimal one obtained in {}``(2)''. In consequence, we choose ``(2)'' as our reference set of optimal SFs, with which the RCG input files given in Supplementary Material are constructed, and the Einstein coefficients are calculated and published in our new database CaDDiAcS \cite{caddiacs}.
Note that this set ``(2)'' is slightly different from the optimal set of Ref.~\cite{lepers2016}, since we have not exactly excluded the same transitions in the two fits.

\subsection{Dynamic dipole polarizabilities}

Using the sum-over-states formula coming from the second-order perturbation theory \cite{lepers2014, li2016, li2017a}, the set of energies and Einstein coefficients obtained above allows for calculating the dynamic dipole polarizabilities (DDPs) of many levels of Er$^+$, in a wide range of wavelengths $\lambda$.
To determine the largest energy and the smallest wavelength for which our data set can be used, we seek to estimate the lowest Er$^+$ energy levels not included in the present calculation. Although none of its levels are known experimentally, the lowest electronic configuration not included in our model is probably $4f^{12}7s$. In Yb$^+$, the corresponding configuration $4f^{14}7s$ appears at 54304.39~cm$^{-1}$ \cite{NIST_ASD}. As expected from the neutral erbium case, the levels of $4f^{12}7s$ certainly play an important role in the DDPs of $4f^{12}6p$ levels, especially for wavelengths close to the $4f^{12}6p$-$4f^{12}7s$ resonances. Similarly, the levels of $4f^{11}6s7s$ are likely to play an important role in the DDPs of $4f^{11}6s6p$ levels. Consequently, the set of data obtained above can be used for energy levels $E$ and vacuum light wavelengths such that $E + hc/\lambda \lesssim 50000$~cm$^{-1}$, and the $6p$-$7s$ transitions should be accounted for with the effective model presented in Refs.~\cite{li2017a, li2018} and used for dysprosium \cite{chalopin2018}.

\begin{figure}
  \includegraphics[width=0.5\textwidth]{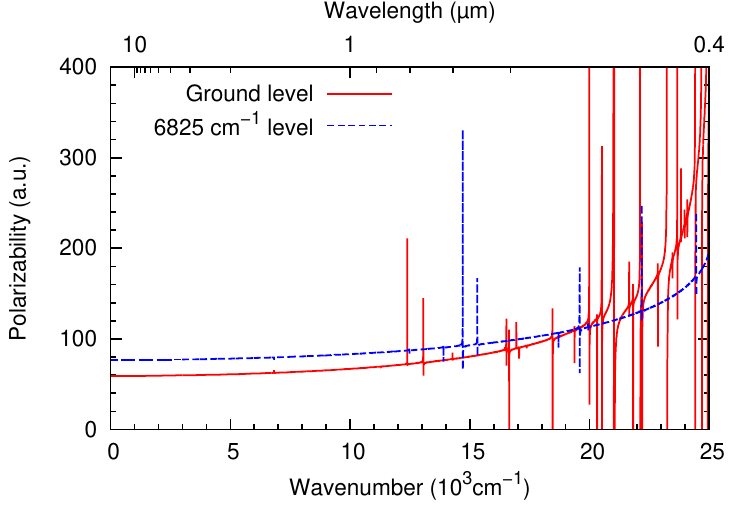}
  \caption{Scalar dynamic dipole polarizability as a function of the wavenumber and the wavelength of the incident light, for the ground level (solid red line) and the excited level at 6824.774~cm$^{-1}$ (dashed blue line) of Er$^+$.
  \label{fig:pola}}
\end{figure}

Among all possible levels, we focus on the ground level $4f^{12}(^3H_6)6s_{1/2} \, (6,1/2)_{13/2}$ and the excited one $4f^{11}6s^2 \,{}^4I^{\text{o}}_{15/2}$ at 6824.774~cm$^{-1}$, whose scalar DDPs, given in Ref.~\cite{li2017a}, Eq.~(7), are plotted on Figure \ref{fig:pola}. The transition between those levels is the equivalent of the clock transition in Yb$^+$ \cite{godun2014, huntemann2016}; but unlike the ytterbium case, that transition is (weakly) allowed in the electric-dipole approximation, with a calculated Einstein coefficient $A_\mathrm{cal} = 16$~s$^{-1}$ (linewidth of 2.6~Hz). Moreover, its vacuum wavelength of 1.465~$\mu$m belongs to the telecommunication band \cite{patscheider2021}.

The static ($\lambda \to \infty$) scalar polarizabilities of the ground and 6825-cm$^{-1}$ levels are respectively 59.2 and 76.8 atomic units (a.u.). The DDP of the ground level increases faster with the wave number and the two DDPs are equal around 19500~cm$^{-1}$. The ground-level DDP also shows many more peaks above 12000~cm$^{-1}$, which is due to the larger number of odd-parity levels compared to even-parity ones in that region of the spectrum (see Fig.~\ref{fig:lev}). 

As for the tensor components of the static polarizabilities, they are equal to $-1.8$ and $-0.9$~a.u.~for the ground and 6825~cm$^{-1}$ levels respectively. Similarly to neutral lanthanides, these small values arise because the polarizabilities of the two levels are mostly due to the isotropic density distribution of the $6s$ electrons, and are thus insensitive to any variation of the electric-field polarization. Another common point with neutrals is that levels belonging to the same manifold have almost equal DDPs. As examples, the (first excited) level $4f^{12}(^3H_6)6s_{1/2} \, (6,1/2)_{11/2}$ at 440.434~cm$^{-1}$ has a scalar (resp.~tensor) static polarizability of 59.2 (resp.~$-1.7$) a.u., and the level $4f^{11}6s^2 \, {}^4I^{\text{o}}_{13/2}$ at 13338.777~cm$^{-1}$ has a scalar (resp.~tensor) static polarizability of 76.5 (resp.~$-0.6$) a.u.


In order to estimate the uncertainty on the DDP, we use the results of Table \ref{tab:opt-sf}. For the levels that we consider, the static static polarizability mostly come from $6s$-$6p$ transitions. Namely, for the ground level, the terms proportional to $\langle 4f^{12}6s |r| 4f^{12}6p \rangle^2$ account for 103~\% of the total value (the other $r_{n\ell,n'\ell'}$ contributions slightly reduce transition dipole moments). For the 6825-cm$^{-1}$ level, the terms proportional to $\langle 4f^{11}6s^2 |r| 4f^{11}6s6p \rangle^2$ are responsible for 97~\% of the total value. In the data set (2) of Table \ref{tab:opt-sf}, the relative uncertainties on $f_1$ and $f_2$ are equal to 5.2 and 5.0~\% respectively. Therefore, we estimate the relative uncertainties to be 10.4 and 10.0~\% for the two levels, which give $59.2 \pm 6.1$ and $76.8 \pm 7.6$ a.u.~respectively.

\section{Conclusion}
\label{sec:ccl}

We have presented a method to perform least-squares fitting of Einstein coefficients by adjusting mono-electronic transition integrals $\langle n\ell |r| n'\ell' \rangle$. This method is implemented in the suite of codes \textit{FitAik} freely available on GitLab \cite{fitaik-gitlab}. The codes are designed to work jointly with either the Dublin \cite{cowan-dublin} or the Kramida \cite{kramida2019} version of the Cowan codes. We have applied our method to the case of Er$^{+}$, for which we have obtained a fair agreement between experimental and calculated Einstein coefficients. The latter can be found on our new database CaDDiAcS \cite{caddiacs}, which currently contains the coefficients of 49122 electric-dipole and 94840 magnetic-dipole transitions.

We think that our least-squares fitting procedure is well suited for atoms with complex structure, such as lanthanides, because a large number of Einstein coefficients are functions of a rather limited number of radial integrals. Therefore, we plan to use our codes to analyze the spectrum of singly-ionized lanthanides, \textit{e. g.} Tm$^{+}$. Moreover, we have already used our codes for neutral atoms, but in a somewhat restricted way. In dysprosium we limited our analysis to the odd-parity configurations $4f^{10}6s6p$ and $4f^{9}5d6s^{2}$; but the Einstein coefficients involving the lowest configuration $4f^{10}6s^{2}$ are sensitive to the configuration interaction with $4f^{9}5d^{2}6s$, which thus will be included in the future \cite{li2016}. The situation is similar to dysprosium, but with configurations $4f^{11}6s6p$, $4f^{10}5d6s^{2}$ and $4f^{10}5d^{2}6s$ \cite{li2017a}. For erbium, accounting for configuration interaction between $4f^{11}5d6s6p$ and other even-parity configuration may surely improve the calculated Einstein coefficients \cite{patscheider2021}.

The major prospect in our work is to improve our method by accounting for the various types of uncertainties. Currently, our code offers the possibility to run several calculations with experimental Einstein coefficients varying randomly within their uncertainty range. In the future, we plan to use weighted least-squares fitting: in the standard deviation, each transition has a weight inversely proportional to its experimental uncertainty. Moreover, we want to provide the user of the CaDDiAcS database with an indication of confidence for each calculated Einstein coefficient \cite{kramida2014}. In this respect, Ref.~\cite{kramida2013} shows, and Figure \ref{fig:ratio-scal} confirms, that the calculated line strength is a good criterion, since the larger the strength, the smaller the discrepancy between theoretical and experimental coefficients. Finally, in addition to the Einstein coefficients and their logarithm, we plan to minimize the standard deviation on the line strength in our least-squares procedure. Preliminary calculations on Er$^+$ do not show strong differences in the resulting optimal scaling factors, but the differences are likely to be large when the range of experimental wavelengths is broad, \textit{e.g.~}in Ref.~\cite{lawler2010}.

\section*{Acknowledgements}

This work is dedicated to Jean-Fran{\c c}ois Wyart who made it possible by sharing with us his invaluable knowledge in atomic spectroscopy. M.L.~acknowledges the financial support of the NeoDip project (ANR-19-CE30-0018-01 from ``Agence Nationale de la Recherche''), and {}``R{\'e}gion Bourgogne Franche Comt{\'e}'' under the project 2018Y.07063 {}``Th{\'e}CUP''. O.D. acknowledges the support by “Investissements d’Avenir” LabEx PALM (ANR-10-LABX-0039-PALM) for the organization of the symposium "in tribute to Jean-Fran{\c c}ois Wyart" held in Orsay, France, on June 22-23, 2022, where this work has been presented. Calculations have been performed using HPC resources from DNUM CCUB (Centre de Calcul de l'Universit\'e de Bourgogne).

\appendix

\section{Gradient and Jacobian for logarithmic standard deviation}
\label{sec:stdev-log}

The gradient vector of the logarithmic standard deviation has components equal to
\begin{equation}
  \frac{\partial \sigma_{\mathrm{log} A}}{\partial f_j} = \frac{2}
    {(N_\mathrm{tr}-N_\mathrm{par}) \ln(10)\, \sigma_{\mathrm{log} A}}
    \sum_t \frac{a_{tj}}{\sqrt{A_{t,\mathrm{cal}}}} 
    \log\left( \frac{A_{t,\mathrm{cal}}}{A_{t,\mathrm{exp}}} \right)
  \label{eq:grad-lg}
\end{equation}
while the Jacobian matrix has elements equal to
\begin{align}
  \frac{\partial^2 \sigma_{\mathrm{log} A}}{\partial f_i \, \partial f_j}
   & = \frac{2}{(N_\mathrm{tr}-N_\mathrm{par}) \ln(10)\, \sigma_{\mathrm{log} A}}
  \nonumber \\
   & \quad \times \sum_t \frac{a_{ti} a_{tj}}{A_{t,\mathrm{cal}}}
    \left[ \frac{2}{\ln 10} 
      - \log \left( \frac{A_{t,\mathrm{cal}}}{A_{t,\mathrm{exp}}} \right) \right]
  \nonumber \\
   & - \frac{4}{(N_\mathrm{tr}-N_\mathrm{par})^2 \ln^2(10)\, \sigma_{\mathrm{log} A}^3} 
  \nonumber \\
   & \quad \times \sum_{t,u} \frac{a_{ti}a_{uj}}
    {\sqrt{A_{t,\mathrm{cal}} A_{u,\mathrm{cal}}}}
    \log\left( \frac{A_{t,\mathrm{cal}}}{A_{t,\mathrm{exp}}} \right)
    \log\left( \frac{A_{u,\mathrm{cal}}}{A_{u,\mathrm{exp}}} \right)
  \label{eq:jacob-lg}
\end{align}
In Eqs.~\eqref{eq:grad-lg} and \eqref{eq:jacob-lg}, the function $\ln$ is the natural (base-$e$) logarithm and $\log$ the base-10 logarithm.

\section{Transitions excluded from the fit}

Table \ref{tab:excl-trs} presents the transitions excluded from the calculation {}``(1)'' of Table \ref{tab:opt-sf}, to give the data set used in {}``(2)'' and {}``(3)''.

\begin{table*}
  \caption{Transitions of calculation {}``(1)'' for which the ratio $A_{\mathrm{cal}} / A_{\mathrm{exp}}$ is smaller than 0.2 or larger than 5. Energies are in cm$^{-1}$ and Einstein coefficients in s$^{-1}$. The notation $(n)$ stands for $\times 10^n$, {}``e'' for {}``even'' and {}``o'' for {}``odd''.
  \label{tab:excl-trs}}
  \begin{center}
  \begin{tabular}{rrr|rrr|rr}
  \hline
   \multicolumn{3}{c|}{Lower level} & \multicolumn{3}{c|}{Upper level} 
    & \multicolumn{2}{c}{Einstein coefficients} \\
    $E_\mathrm{exp}$ & Parity & $J$ & $E_\mathrm{exp}$ & Parity & $J$
     & $A_\mathrm{cal}$ & $A_\mathrm{exp}$ \\
  \hline
    5133 & e &  $9/2$ & 35928 & o &  $7/2$ & 7.072(4) & 4.500(5) \\
    5404 & e &  $7/2$ & 35928 & o &  $7/2$ & 5.924(6) & 1.060(6) \\
    7150 & e & $11/2$ & 37147 & o & $11/2$ & 4.246(6) & 5.900(7) \\
    7195 & e &  $9/2$ & 37147 & o & $11/2$ & 1.538(6) & 2.460(7) \\
    5133 & e &  $9/2$ & 34342 & o & $11/2$ & 2.787(4) & 1.000(6) \\
    5404 & e &  $7/2$ & 34196 & o &  $9/2$ & 8.568(5) & 1.430(5) \\
    5133 & e &  $9/2$ & 33660 & o & $11/2$ & 7.534(6) & 1.360(6) \\
    5404 & e &  $7/2$ & 33539 & o &  $7/2$ & 1.809(6) & 1.200(7) \\
    5133 & e &  $9/2$ & 32896 & o &  $9/2$ & 9.673(2) & 2.090(6) \\
    5133 & e &  $9/2$ & 32811 & o & $11/2$ & 9.307(3) & 7.900(4) \\
    5404 & e &  $7/2$ & 32896 & o &  $9/2$ & 8.786(6) & 1.290(6) \\
    7195 & e &  $9/2$ & 34674 & o &  $9/2$ & 1.197(5) & 6.100(5) \\
    7150 & e & $11/2$ & 33028 & o &  $9/2$ & 2.818(7) & 5.300(6) \\
    7150 & e & $11/2$ & 32073 & o & $11/2$ & 3.278(4) & 1.930(5) \\
   11043 & e &  $9/2$ & 35928 & o &  $7/2$ & 2.838(5) & 1.530(6) \\
   16936 & o & $19/2$ & 40547 & e & $17/2$ & 1.305(3) & 5.100(5) \\
    7195 & e &  $9/2$ & 30318 & o &  $9/2$ & 3.064(2) & 2.700(4) \\
   10894 & e &  $7/2$ & 33539 & o &  $7/2$ & 8.512(3) & 1.510(5) \\
   11043 & e &  $9/2$ & 33660 & o & $11/2$ & 2.363(3) & 2.500(4) \\
   17064 & o & $11/2$ & 39083 & e & $13/2$ & 1.700(5) & 1.370(6) \\
   10894 & e &  $7/2$ & 29784 & o &  $9/2$ & 2.714(3) & 1.230(5) \\
   18889 & e & $11/2$ & 32528 & o & $13/2$ & 1.132(5) & 5.800(5) \\
  \hline
  \end{tabular}
  \end{center}
\end{table*}

\begin{table*}
  \caption{Pairs of close odd-parity levels having the same $J$ values and experimental energies separated by less than 100~cm$^{-1}$.}
  \label{tab:lev-cls}
  \begin{center}
  \begin{tabular}{c|cc|cc}
  \hline
$J$ &$E_\mathrm{exp}$&$E_\mathrm{cal}$&$g_\mathrm{exp}$&$g_\mathrm{cal}$ \\ \hline
11/2&30122           &30138           &1.125           &1.136  \\
    &30157           &30106           &1.070           &1.077  \\ \hline
9/2 &33566           &33553           &1.100           &1.078  \\
    &33650           &33646           &1.030           &1.042  \\ \hline
11/2&33660           &33681           &1.175           &1.194  \\
    &33721           &33910           &1.190           &1.170  \\ \hline
7/2 &34148           &34148           &1.255           &1.173  \\
    &34203           &34228           &1.200           &1.195  \\ \hline
7/2 &36738           &37657           &1.086           &1.080  \\
    &36825           &36840           &1.160           &1.156  \\ \hline
9/2 &37039           &37067           &1.145           &1.123  \\ 
    &37110           &37222           &1.065           &1.040  \\ \hline
11/2&37098           &37130           &1.250           &1.110  \\
    &37147           &37201           &1.150           &1.195  \\ \hline
5/2 &38580           &38645           &0.935           &0.947  \\
    &38617           &38776           &0.990           &1.088  \\ \hline 
7/2 &39277           &39254           &1.215           &1.031  \\
    &39304           &39348           &1.025           &1.192  \\ \hline
  \end{tabular}
  \end{center}
\end{table*}

\bibliographystyle{elsarticle-num}


\end{document}